# Mathematical Modeling of Isolated Wind-Diesel-Solar Photo Voltaic Hybrid Power System for Load Frequency Control

Bharat Pariyar[1] Raju Wagle[2]

[1]Narvik University College, 8514 Narvik, Norway

[2]Pokhara University, School of Engineering, Lekhnath 12, Kaski, Nepal

**Abstract:** *This research presents the mathematical model of an isolated wind-diesel-solar PV hybrid power system with conventional proportional-plus-integral controllers for load frequency control (LFC). In order to enhance the reliability of the power supply, renewable sources such as wind and solar energy are integrated with diesel electric power generation system to supply the power for isolated loads. Isolated hybrid power system is designed to minimize the mismatch between supply and demand. Due to the unstable generation of power from wind, solar PV sources, and frequent change in load, there exist fluctuations of power generation and hence fluctuations also occur in system frequency and voltage. Conventional PI controllers are used for the load frequency control of the system to make the frequency deviation to an acceptable range. In this paper the complete mathematical modeling of system consisting of a wind turbine induction generator unit, a diesel engine synchronous alternator unit and solar photovoltaic (PV) panels with maximum power tracking converter is presented.*

## 1. Introduction

There are several important reasons that make renewable energy important for the future of our society. By using renewable energy instead of fossil fuels, we can significantly decrease the current levels of greenhouse gas emissions. Renewable energy such as solar energy and wind energy are endless resources as compared to the conventional fossil fuels. Owing to high demand of electricity in modern society and wide gap between supply and demand it is very difficult to fulfill the need of electricity only with the conventional sources. Therefore, renewable energy sources such as solar; wind, biomass etc. are emerging in today world. However both wind and solar energy are intermittent in nature which not only changes the generation but also affect the system voltage and frequency.[1,2] Hence, solar PV and wind power generations are integrated with diesel system in order to supply reliable, secure and economical power to the isolated loads [3-5].

Due to the fluctuations in frequency, system becomes unstable and hence effective controllers are required for maintaining the system frequency to an acceptable range either by maintaining the load fluctuation or by controlling the generation. There are different control strategies to control the mismatch between load and generation [6-10]. Different strategies are priority switched load control [5], fly wheel [6], dump load control [9], battery energy storage [10] and superconducting magnetic energy storage [8]. These strategies are expensive and they have their own limitations. In an isolated wind-diesel-Solar PV hybrid power system, load frequency control (LFC) scheme is used. This strategy is used to obtain the acceptable frequency and hence it is useful for maintaining the system's performance [11].

In this research, detailed analytical study for an isolated wind-diesel-solar PV hybrid power system with complete mathematical modeling under transient conditions by considering a small signal transfer function model is done. The configuration of an isolated wind-diesel-solar PV hybrid power system is shown in figure 1.

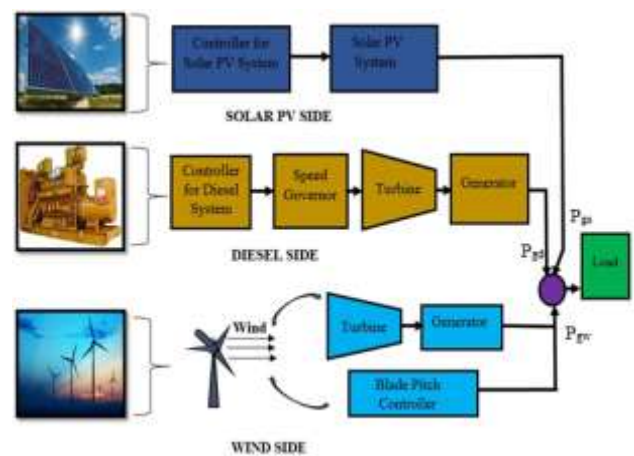

**Figure 1:** Configuration of an isolated Wind-Diesel-Solar PV hybrid power system

## 2. Mathematical Modeling of System

### 2.1 Mathematical modeling of solar PV system

PV panel model consists of solar cells and each panel is made from the different series-parallel combination of these solar cells. Every solar cell acts as p-n diode and hence current passes from one side to the other side [12]. The equivalent circuit of a solar cell is shown in figure 2.

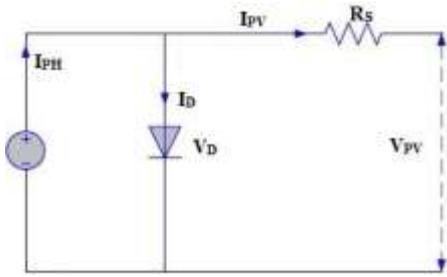

**Figure 2:** Equivalent circuit of solar cell

$I_{ph}$ is used as the reference current and $R_s$ is equivalent to the total resistance. The equations are as follows [12]

$$I_{pv} = I_{ph} - I_{SAT}\left(\frac{q(V_{PV}+I_{PV}R_s)}{(AKT-1)}\right) \quad (1)$$

$$I_{ph} = \left(\frac{\lambda}{1000}\right)[I_{sc} + K_I(T-25)] \quad (2)$$

we use MPPT controller for regulating PV output voltage and boost converter to achieve AC regulating voltage. The relevant circuit is shown in figure 3.[12]

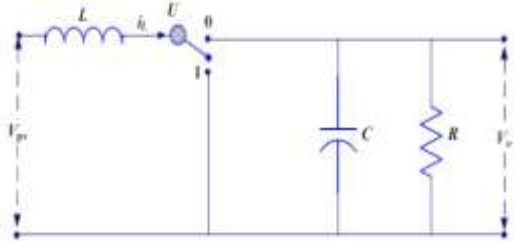

**Figure 3:** Boost Converter equivalent circuit

U switch shown in this figure 3 is composed of an IGBT and Diode. When U=0, Diode is ON and IGBT is OFF and vice versa.

Switch operation can be divided into two different time periods [13]. First period occurs when the switch is turned ON, it means $0 \leq t \leq t_{ON}$, the relevant equations are as follows:

$$L\frac{di_L}{dt} = V_{PV} \quad (3)$$

$$C\frac{dv_o}{dt} + \frac{v_o}{R} = 0 \quad (4)$$

Now another time period belongs to turned OFF switch, it means $t_{ON} \leq t \leq T_S$

$$L\frac{di_L}{dt} + V_O = V_{PV} \quad (5)$$

$$i_L - C\frac{dv_o}{dt} + \frac{v_o}{R} = 0 \quad (6)$$

$T_S$ represents a switching time period.

First, the transfer function model of MPPT, filter, inverter and PV panel are derived [14, 15] and hence the transfer function model of the PV panel can be find out from the above equations 1 and 2.

MPPT is done by the boost converter and we have to consider the boost converter's ON state and OFF state mode of operations [14], which are given by the equations 3, 4, 5 and 6. Combining all these equations, transfer function model of the PV Panel is obtained. It is given in the equation 7.

$$G_{BC} = \frac{-18S+900}{S^2+100S+50} \quad (7)$$

The change in temperature and irradiation is the step input in the PV panel.

From the mathematical model of the solar photo voltaic, the transfer function block diagram of the solar PV generating system is developed as shown in Figure 4.

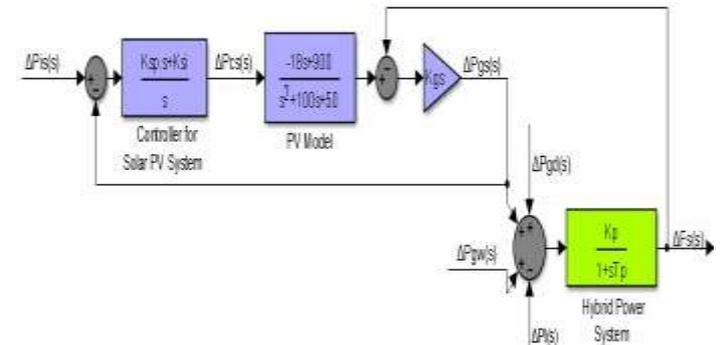

**Figure 4:** Transfer function model of solar PV system

### 2.2 Mathematical modeling of Diesel System

The conversion of fuel energy (diesel or bio-diesel) into mechanical energy and then into electric energy is due to the act of diesel generator sets [16]. Imbalance occurs between the real power generation and the load demand (plus losses) which causes kinetic energy of rotation to be either added to or taken from the generating units (generator shaft either speed up or slow down). This varies the frequency of the system [17], and the governor maintains the balance between the input and output by changing the turbine output and the PI controller uses a system frequency deviation of the power system as a feedback input.

The transfer function of the mechanical speed-governing system in diesel unit can be written in partial fraction form as in equation 8.

$$\frac{K_d(1+sT_{d1})}{(1+sT_{d2})(1+sT_{d3})} = \frac{K_1}{(1+sT_{d2})} + \frac{K_2}{(1+sT_{d3})} \quad (8)$$

Where

$$K_1 = \frac{K_d(1+sT_{d1})}{(1+sT_{d3})} = \frac{K_d(T_{d2}-T_{d1})}{(T_{d2}-T_{d3})} \text{ at } s = -\frac{1}{T_{d2}} \quad (9)$$

And

$$K_2 = \frac{K_d(1+sT_{d1})}{(1+sT_{d2})} = \frac{K_d(T_{d3}-T_{d1})}{(T_{d3}-T_{d2})} \text{ at } s = -\frac{1}{T_{d3}} \quad (10)$$

Td1, Td2 and Td3 are the time constants of the speed governing mechanism and $K_d$ is the part of power supplied by diesel power generation to the load. Equation (8) can be written in terms of the canonical state variables $\Delta X_{ED11}$ and $\Delta X_{ED21}$,

$$\frac{K_d(1+sT_{d1})}{(1+sT_{d2})(1+sT_{d3})}\left[\Delta P_{cd}(s) - \frac{1}{R_d}\Delta F_s(s)\right] = \Delta X_{ED11}(s) + X_{ED21}(s)$$

(11)

Where Rd is the speed regulation due to the governor speed action and from equation (8) and equation (11), we get

$$\Delta X_{ED11}(s) = \frac{K1}{(1+sTd2)} \left[ \Delta Pcd(s) - \frac{1}{Rd}\Delta Fs(s) \right] \quad (12)$$

And

$$\Delta X_{ED21}(s) = \frac{K2}{(1+sTd3)} \left[ \Delta Pcd(s) - \frac{1}{Rd}\Delta Fs(s) \right] \quad (13)$$

Therefore, the state differential equations of the mechanical speed governing mechanism are written in equations (14) and (15).

$$\frac{d}{dt}\Delta X_{ED11} = -\frac{1}{Td2}\Delta X_{ED11} - \frac{Kd(Td2-Td1)}{Rd.Td2(Td2-Td3)}\Delta Fs + \frac{Kd(Td2-Td1)}{Td2(Td2-Td3)}\Delta Pcd \quad (14)$$

$$\frac{d}{dt}\Delta X_{ED21} = -\frac{1}{Td3}\Delta X_{ED21} - \frac{Kd(Td3-Td1)}{Rd.Td3(Td3-Td2)}\Delta Fs + \frac{Kd(Td3-Td1)}{Td3(Td3-Td2)}\Delta Pcd \quad (15)$$

The transfer function equation for the change in diesel power generation $\Delta Pgd$, can be written in terms of the state variables as

$$\Delta P_{gd}(s) = \frac{1}{(1+sTd4)} \left[ \Delta X_{ED11}(s) + \Delta X_{ED21}(s) \right] \quad (16)$$

$$\frac{d}{dt}\Delta P_{gd} = -\frac{1}{Td4}\Delta P_{gd} + \frac{1}{Td4}\Delta X_{ED11} + \frac{1}{Td4}\Delta X_{ED21} \quad (17)$$

The transfer function block diagram of the diesel-generating unit is shown in figure 5.

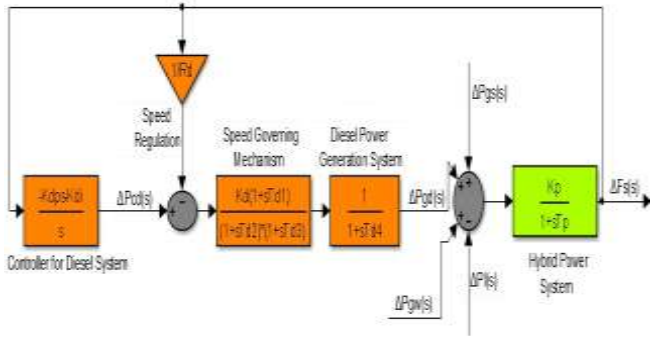

**Figure 5:** Transfer function model of diesel system

## 2.3 Mathematical Modeling of Wind System

In the wind-turbine generating unit, blade pitch controller constantly maintains the wind power generation. The intermittent wind power may affect the power quality of an isolated wind-diesel-Solar PV hybrid power system and the deviations in generating power and frequency fluctuations are eliminated by blade pitch control mechanism, which continuously monitors the wind turbine speed and acts accordingly in an active feedback control system added to the turbine.

The transfer function equation for the wind generation system is,

$$\Delta Ft(s) = \frac{1}{1+sTw}[-\Delta Pgw(s) + \Delta Piw(s) + \Delta Pcw(s) + Ktp \Delta Ft(s)] \quad (18)$$

and

$$\Delta Pgw(s) = Kig [\Delta Ft(s) - \Delta Fs(s)] \quad (19)$$

Where
Tw is the time constant of the wind-turbine power generation system in sec.
Kig is a function of slip and is the part of power supplied by wind-power generation to load.
Ktp is the co-efficient that depends on the slope and curve of the wind turbine [18]

From Equation (18) and Equation (19) the state differential equation can be written as

$$\frac{d}{dt}\Delta Ft(s) = -\frac{(1+Kig-Ktp)}{Tw}\Delta Ft + \frac{Kig}{Tw}\Delta Fs + \frac{1}{Tw}\Delta Piw + \frac{1}{Tw}\Delta Pcw \quad (20)$$

The real power load change $\Delta Pl$ or change in wind power generation $\Delta Pgw$ experienced by the hybrid system deviates the power generation from a specified level and the power generation of the hybrid system can be maintained by the diesel engine controller by changing its power generation by an amount $\Delta Pgd$. The net surplus power $\Delta P_I$ will be absorbed by the system either by increasing the kinetic energy of the system or by increased load consumption.

The surplus power is,
$$\Delta P_I = [\Delta Pgd + \Delta Pgw - \Delta Pl] \quad (21)$$

The transfer function equation of the system subjected to change in real power load or input wind power can be written as in equation (22).

$$\Delta Fs = \frac{Kp}{1+sTw}[\Delta Pgd(s) + \Delta Pgw(s) - \Delta Pl(s)] \quad (22)$$

Where

$$Kp = \frac{1}{D}$$

$$D = \frac{\partial Pl}{\partial f}$$

$$Tp = \frac{2H}{Fs.D}$$

H = P.U. Inertia constant
Fs = Nominal system frequency
D = Damping coefficient

The state differential equation is represented by equation (23).

$$\frac{d}{dt}\Delta Fs = -\frac{1+Kig.Kp}{Tp}\Delta Fs + \frac{Kp}{Tp}\Delta Pgd + \frac{Kig.Kp}{Tp}\Delta Ft - \frac{Kp}{Tp}\Delta Pl \quad (23)$$

The combined transfer function of different blocks of the blade pitch control mechanism is given in equation (24).

$$\left[\frac{Kpc.Kp3}{1+sTp3}\right]\left[\frac{Kp2}{1+sTp2}\right]\left[\frac{Kp1(1+sTp1)}{(1+s)}\right]\Delta Pcu(s) = \Delta Pcw(s) \quad (24)$$

Where
Tp1, Tp2 are the time constants of the hydraulic blade pitch actuator in sec
Tp3 is the time constant of the data fit pitch response unit
Kp1 and Kp2 are gain constants of the hydraulic pitch actuator
Kp3 is the gain constant of the data fit pitch response unit
Kpc is the blade characteristic constant

Equation (24) can be written as

$$\left[\frac{Kpc.Kp3}{1+sTp3}\right][Kp1\{Tp1+\frac{(1-Tp1)}{(1+s)}\}][\frac{Kp2}{1+sTp2}]\Delta Pcu(s) = \Delta Pcw(s) \quad (25)$$

Equation (25) can be expressed in terms of intermediate state variables as

$$\Delta Pcw(s) = \left[\frac{Kpc.Kp3}{1+sTp3}\right][Kp1.\Delta P_{C1}(s) + Kp1.Tp1.\Delta P_{C2}(s)] \quad (26)$$

$$\Delta P_{C1}(s) = \frac{(1-Tp1)}{(1+s)}\Delta P_{C2}(s) \quad (27)$$

$$\Delta P_{C2}(s) = \frac{Kp2}{1+sTp2}\Delta Pcu(s) \quad (28)$$

The state differential equations for the transfer function Equation (26), Equation (27) and Equation (28) are given by equations (29), (30) and (31) respectively.

$$\frac{d}{dt}\Delta Pcw = -\frac{1}{Tp3}\Delta Pcw + \frac{Kpc.Kp3.Kp1}{Tp1}\Delta P_{C1} + \frac{Kpc.Kp3.Kp1}{Tp3}\Delta P_{C2} \quad (29)$$

$$\frac{d}{dt}\Delta P_{C1} = -\Delta P_{C1} + (1-Tp1)\Delta P_{C2} \quad (30)$$

$$\frac{d}{dt}\Delta P_{C2} = -\frac{1}{Tp2}\Delta P_{C2} + \frac{Kp2}{Tp2}\Delta Pcu \quad (31)$$

The transfer function block diagram of the wind-turbine generation system with blade pitch controller is shown in figure 6.

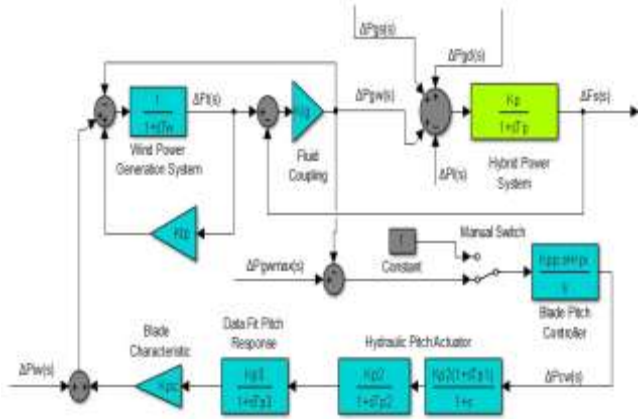

Figure 6: Transfer function model of wind system

### 2.4 Overall System Modeling

The transfer function block diagram of an isolated wind-diesel-Solar PV hybrid power system is shown in figure 7. PI controllers are included in the transfer function block diagram model of the hybrid system for load frequency control. The input power to the wind side and solar PV side are not controllable. There is small real power mismatch and the system dynamics may be described by linear differential equations [19-22]. The functions of the controllers are used to eliminate the mismatch created either by the small real power load change or due to a change in input power. Conventional PI controllers are designed for the load frequency control of an isolated wind-diesel-solar PV hybrid power system. PI controller for a governor in diesel side, blade pitch controller in wind side and PI controller in solar PV system are designed individually for performance improvement of an isolated wind-diesel-solar PV hybrid power system, which is shown in the figure 7. The input power to the renewable sources of power generation is fluctuating, particularly in case of wind by nature and in case of solar PV system due to uncertainty in the availability of solar power. In figure 7, ΔFs and ΔFt represent, respectively, deviations in system frequency (60 Hz) and speed of the wind-turbine induction generator. ΔPgd, ΔPgw and ΔPgs represent deviations in diesel, wind and PV power generation, respectively.

The dynamics of the wind power generating unit is described by a first order system and a higher order model [23-24] and the continuous time dynamic behavior of the load frequency control system is modelled by a set of state space differential equations of the form as in equation (32).

$$\dot{X} = AX + Bu + \Gamma p \quad (32)$$

Where
$\dot{X}$, u and p are the state, control and disturbance vectors, respectively.
A, B and Γ are real constant system matrices of appropriate dimensions.

The elements of the matrices in (32) along with the principal data of the system under study are given in [20].

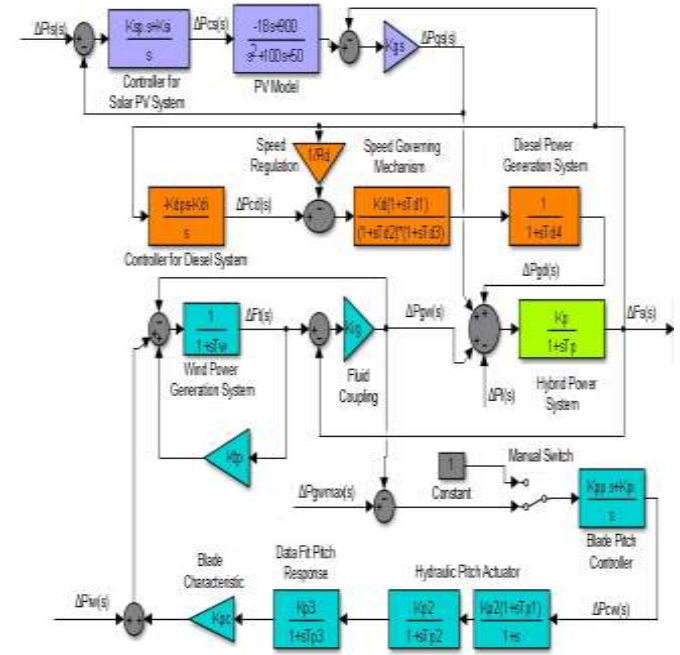

**Figure 7:** Transfer function block diagram for an isolated Wind-Diesel-Solar PV hybrid power system with controllers

## 3. Conventional Pi Controller for LFC

### 3.1 Introduction of Pi Controller

PI controller is most widely used for load frequency control schemes. The importance of such controller is for reducing the steady-state error to zero. The block diagram of PI controller is shown in figure 8. [11].

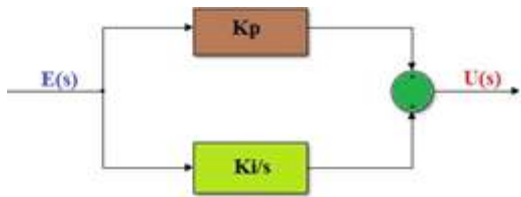

**Figure 8:** Block diagram of PI controller

Mathematically, transfer function of PI controller can be represented as,

$$\frac{U(s)}{E(s)} = K_P + \frac{K_i}{s} \quad (33)$$

### 3.2 Modeling of PI Controller

The task of this thesis is to investigate the problem in the control of system frequency for an isolated hybrid power system using PI controller designed here. For PI controller type load frequency controller of proportional plus integral type in isolated wind-diesel-Solar PV hybrid power system in case of continuous case, to achieve zero steady state error in frequency, can be obtained by augmenting the state vector in (32). $x_{n+1}$ and $x_{n+2}$ are two additional state variables which are defined in equations (34) and (35). [23].

$$X_{n+1} = \int \Delta F_s \, dt \quad (34)$$
$$X_{n+2} = \int \Delta F_t \, dt \quad (35)$$

Therefore, the additional state differential equations can be written as

$$\dot{X}_{n+1} = \Delta F_s \quad (36)$$
$$\dot{X}_{n+2} = \Delta F_t \quad (37)$$

Equations (36) and (37) can be written in matrix form as

$$\begin{bmatrix} \dot{X}_{n+1} \\ \dot{X}_{n+2} \end{bmatrix} = A_1 X \quad (38)$$

Now the state vector in equation (3.32) is modified by including the state variables defined in equations (34) and equation (35).

The augmented set of differential equations is shown in equation (39).

$$\hat{\dot{x}} = \begin{bmatrix} A & O_1 \\ A_1 & O_2 \end{bmatrix} \hat{X} + \begin{bmatrix} B \\ O_3 \end{bmatrix} u + \begin{bmatrix} \Gamma \\ O_4 \end{bmatrix} p \quad (39)$$

Where $O_1, O_2, O_3$ and $O_4$ are null matrices of appropriate dimensions and the control vector 'u' can be expressed in terms of the augmented state vector as in equation (40)

$$u = H \hat{X} \quad (40)$$

Where
$H =$
$$\begin{bmatrix} -K_{dp} & 0 & 0 & 0 & 0 & 0 & 0 & 0 & -K_{di} & 0 \\ K_{ig}K_{pp} & 0 & 0 & 0 & -K_{ig}K_{pp} & 0 & 0 & 0 & K_{ig}K_{pi} & -K_{ig}K_{pi} \end{bmatrix}$$
$$(41)$$

Now the final augmented set of differential equations can be written as

$$\hat{\dot{X}} = \hat{A}\hat{X} + \acute{\Gamma} p \quad (42)$$

Where

$$\hat{A} = \begin{bmatrix} A & O_1 \\ A_1 & O_2 \end{bmatrix} + \begin{bmatrix} B \\ O_3 \end{bmatrix} H \quad (43)$$

And

$$\acute{\Gamma} = \begin{bmatrix} \Gamma \\ O_4 \end{bmatrix} \quad (44)$$

## 4. Conclusion

A complete mathematical model of an isolated wind-diesel-solar photo voltaic system on the basis of small signal transfer function model together with controller design of the load frequency controller is modeled in the paper. This model can be further used in load frequency control with various optimization techniques.

## Author Profile

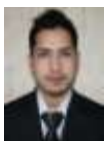
**Bharat Pariyar** received his Bachelor Degree in Electrical Engineering from Kathmandu Engineering College (Tribhuvan University) and Masters Degree in Electrical Engineering from Narvik University College Norway in the year 2015. His research interest includes Hydropower, Power system operation and control, Energy system management and Load frequency control in hybrid power system. He is a member of Nepal Engineering Council (NEC) and is working as an Electrical Engineer in Oslo, Norway.

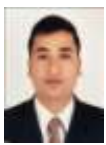
**Raju Wagle** received his Bachelor degree in Electrical Engineering from Institute of Engineering, Pulchowk campus in the year 2010 and his master's degree in Electrical Engineering from Narvik University College, Norway in 2015. His research interests are not only limited to renewable energies, power system integration, operation and control of grid and isolated power system, power electronics and many more. He is the permanent faculty of Electrical and Electronics Department, School of Engineering, Pokhara University. Besides he is also engaged in an Energy related sector in Nepal.

## Appendix

List of Symbols And Abbreviations

### Symbols

$K_{pc}$    Blade characteristic constant, pu kW / deg.
$K_{pi}$    Blade pitch controller integral gain
$K_{pp}$    Blade pitch controller proportional gain
$K_{si}$    Integral gain of solar PV controller
$K_{sp}$    Proportional gain of solar PV controller
$\Delta P_{cw}$    Change in blade angle position
$\Delta P_{cd}$    Change in diesel engine speed changer position
$\Delta P_{gd}$    Change in diesel power generation, pu kW
$\Delta P_{gs}$    Change in solar PV generation, pu kW
$\Delta P_{is}$    Change in input solar PV power
$\Delta P_{iw}$    Change in input wind power due to change of wind velocity
$\Delta PI$    Change in net surplus power absorbed by the system
$\Delta P_l$    Change in real power load
$\Delta P_{gw}$    Change in wind power generation, pu kW
$\Delta F_t$    Change in wind turbine speed
$K_{tp}$    Coefficient that depends on the slope of $C_p$, $\lambda$ curves of the wind turbine
B    Control Matrix
U    Control vector
D    Damping coefficient
$K_d$    Derivative gain
$\Gamma$    Disturbance Matrix
P    Disturbance vector
$K_{p1}$, $K_{p2}$ Gain constant of hydraulic blade pitch actuator
$K_{p3}$    Gain constant of the data fit pitch response unit
$\Delta XED$    Incremental change in governor valve position
$K_i$    Integral gain
$K_{di}$    Diesel controller integral gain
$K_{dp}$    Diesel controller proportional gain
$F_s$    Nominal system frequency
Ɩŋ    Performance index
$K_p$    Power system gain
$T_p$    Power system time constant in sec.
$KP$    Proportional gain
H    P.U. Inertia constant
$R_d$    Speed regulation to the governor action in Hz / pu kW
X    State vector
$\Delta F_s$    System frequency deviation, Hz
A    System Matrix
$T_{d1}$, $T_{d2}$, $T_{d3}$ Time constant of diesel engine speed governing mechanism in sec.
$T_{p1}$, $T_{p2}$ Time constant of the hydraulic blade pitch actuator in sec.
$K_d$    The part of power supplied by diesel power generation to the load
$K_{gs}$    The part of power supplied by Solar PV to the load
$K_{ig}$    The part of power supplied by wind to the load
$T_{d4}$    Time constant of diesel power generation in sec.
$T_{p3}$    Time constant of the data fit pitch response unit in sec.
$T_w$    Time constant of the wind-turbine in sec.

ISC    Short circuit current (A)
IPV    Photovoltaic current (A)
IPH    Photo current (A)
ISAT   Saturation current (A)
ID     Diode current (A)
q      Charge of electron = $1.602*10^{-19}$ (C)
RS     Resistance
λ      Solar Irradiance ( $w/m^2$ )
T      Temperature of solar array ( º C)
A      Diode quality factor
KI     short- circuit current
VPV (min)   Minimum output voltage
VPV (max)   Maximum output voltage of PV
vo (max)    Maximum output voltage
fsw    Switching frequency
Io     Load current
L      Inductance
C      Capacitance

**Abbreviations**
LFC    Load Frequency Control
BPC    Blade Pitch Control
PV     Photovoltaic
PI     Proportional Integral
GA     Genetic Algorithm
HVDC   High Voltage Direct Current

**Table 1:** Rating of proposed hybrid system

| Generation Capacity | | |
|---|---|---|
| Wind Generation(kW) | Diesel Generation(kW) | PV Generation(kW) |
| 150 | 150 | 60 |
| Load of the system = (150 +100 + 50) = 300 kW | | |

**Table 2:** Values of system parameters

| System Parameters |
|---|
| Td1 = 1 s ;    Td2 = 2 s ;    Td3 = 0.025 s ;    Td4 = 3 s |
| Rd = 5 Hz/pu KW ;    Tw = 4 s ;    Kpc = 0.08 pu Kw/deg |
| Kp1 = 1.25 ;    Kp2 = 1 ;    Kp3 = 1.4 ; |
| Kp1*Kp2*Kp3 = 1.75 deg/pu KW ;    F = 60 Hz |
| Tp1 = 0.6 s ;    Tp2 = 0.041 s ;    Tp3 = 1 s |
| Kig = 0.9969 pu kW/Hz ;    Kd = 0.3333 pu kW/Hz |
| Ktp = 0.003333 pu kW/Hz    Kgs = 0.20 pu kW/Hz |
| Kp = 72 ;    Tp = 14.4 sec |